\documentclass[twocolumn,aps,reprint,amsmath,amssymb,graphicx]{revtex4-2}
\usepackage{graphicx}
\usepackage{amsfonts}
\usepackage{amsmath}
\usepackage{hyperref}
\usepackage{subcaption}
\usepackage{todonotes}  
\usepackage{notes2bib}
\usepackage{braket} 
\hypersetup{
	colorlinks,  
	citecolor=black, 
	filecolor=black, 
	linkcolor=black,
	urlcolor=black
      }    

\usepackage{todonotes}

\newcommand{\cf}{CF$_3^+$}

\newcommand{\co}{COCH$_3^+$}
\newcommand{\cfn}{CF$_3$}     
\newcommand{\con}{COCH$_3$}   

\makeatletter
\newcommand{\customlabel}[2]{%
   \protected@write \@auxout {}{\string \newlabel {#1}{{#2}{\thepage}{#2}{#1}{}} }%
   \hypertarget{#1}{}
}
\makeatother

\begin{document}
\title{Direct Observation of Entangled Electronic-Nuclear Wave Packets}
\author{G\" onen\c c Mo\u gol}
\affiliation{Department of Physics and Astronomy, Stony Brook University, Stony Brook NY 11794-3800, USA}
\author{Brian Kaufman}
\affiliation{Department of Physics and Astronomy, Stony Brook University, Stony Brook NY 11794-3800, USA}
\author{Chuan Cheng}
\affiliation{ Stanford PULSE Institute, SLAC National Accelerator Laboratory, Menlo Park, California 94025, USA}
\affiliation{Department of Physics and Astronomy, Stony Brook University, Stony Brook NY 11794-3800, USA}

\author{Itzik Ben-Itzhak}
\affiliation{James R. Mcdonald Laboratory, Kansas State University, Manhattan KS 66506-2604, USA}

\author{Thomas Weinacht}
\email{Author to whom correspondence should be addressed: thomas.weinacht@stonybrook.edu}
\affiliation{Department of Physics and Astronomy, Stony Brook University, Stony Brook NY 11794-3800, USA}

\begin{abstract}
    We present momentum resolved covariance measurements of
    entangled electronic-nuclear wave packets created and
    probed with octave spanning phaselocked ultrafast
    pulses.  We launch vibrational wave packets on multiple
    electronic states via multi-photon absorption, and probe
    these wave packets via strong field double ionization using a
    second phaselocked pulse. Momentum resolved
    covariance mapping of the fragment ions highlights the nuclear motion, while
    measurements of the yield as a function of the relative phase between pump and probe pulses highlight the electronic coherence.  The combined
    measurements allow us to directly visualize the
    entanglement between the electronic and nuclear degrees of
    freedom and follow the evolution of the complete
    wavefunction. 
\end{abstract}  

\maketitle
\section{Introduction}

The interplay between electronic and nuclear dynamics is at
the forefront of attosecond science and of fundamental
importance for understanding energy flow at the molecular
level \cite{corkum2007attosecond,palacios2020quantum,koll2022,Vrakking_2022,vrakking21}. Of
particular interest is the role that electronic coherences
play in driving basic photophysical processes such as
photoisomerization, light harvesting, and how these
coherences are affected by nuclear dynamics \cite{duan2017nature,maiuri2018coherent}. While electronic
coherences in atoms can last for several
nanoseconds and have been studied extensively 
\cite{PhysRevLett.92.133004,PhysRevLett.80.5508,PhysRevLett.64.2007}, measurements
of coherences in large molecules have been limited to a few
femtoseconds \cite{arnold2017electronic, Vibok,franco2008femtosecond,scheidegger2022search,hwang2004electronic,kamisaka2006ultrafast,PhysRevLett.118.083001,PhysRevA.92.040502}, or in special cases a few tens of femtoseconds
\cite{Despre2018}. 

The rapid loss of electronic coherence is driven by the coupling between the electrons and the nuclei. This can be seen by writing the full molecular wave function after photo-excitation in a Born-Huang (Oppenheimer) expansion:

  \begin{multline}
      \Psi(\boldsymbol{r},\boldsymbol{R},t)=a_1(t)
      \chi_1(\boldsymbol{R},t)
      \psi_1(\boldsymbol{r};\boldsymbol{R})e^{-i
        \omega_1(\boldsymbol{R})t}+\\ a_2(t)
      \chi_2(\boldsymbol{R},t)
      \psi_2(\boldsymbol{r};\boldsymbol{R})e^{-i
        \omega_2(\boldsymbol{R})t}+\cdots \; ,
      \label{eq:entangled_wfn}
  \end{multline}
where $\psi_i$ represents the $i$$^\mathrm{th}$ electronic state of the molecule and $\chi_i$ represents the nuclear wavefunction on the $i$$^\mathrm{th}$ electronic state. The frequencies in the
exponent $\omega_i$ are defined by the
formula $\omega_{i}(\boldsymbol R) =
V_{i}(\boldsymbol{R})/\hbar$, where $V_{i}(\boldsymbol{R})$
is the potential energy surface of state $i$ as a function
of the nuclear coordinate $\boldsymbol{R}$.  The coefficients
$a_i$ can depend on time because internal conversion or intersystem crossing can lead to a changes in the population of a given electronic state; however, this is not relevant in this experiment.

The wave packet described by Equation
\ref{eq:entangled_wfn} can be created by multiphoton
absorption using a very short ($\sim$2 cycle) pulse \cite{kaufman2022numerical}, such that the
excitation is dominated by a single sub-cycle of the
pulse. In the case of $n$--photon absorption, the effective
bandwidth of the $n^\mathrm{th}$ order Rabi frequency is
$\sqrt{n}$ times the spectral bandwidth, allowing for
excitation of multiple electronic states for any given
photon order.  When driving $n$--photon absorption, it is
natural to expect $n\pm1$ order excitation of nearby states
\cite{kaufman2022numerical,lunden2014model,brian-coherence}.

Here we focus on the special case of
\autoref{eq:entangled_wfn} for two excited states which are separated by one photon order (e.g., 4-- and 5--photon absorption). The off
diagonal term of the density matrix, which expresses the
electronic coherence is given by:

\begin{equation}
    \rho_{12}(\boldsymbol R,t)=a_1 \chi_1(\boldsymbol{R},t)
    a_2^* \chi_2^*(\boldsymbol{R},t) e^{i(\omega_2(\boldsymbol{R})-\omega_1(\boldsymbol{R}))t}
    \label{eq:coherence}
\end{equation}

In our experiment we are considering an $n$--photon absorption to
the state 1, and $(n+1)$--photon absorption to state 2 by the pump pulse, which we can write
as $E_{\text{pu}}(t) = E_{0}(t) \cos(\omega_{0}t)$. Subsequently,
a probe pulse doubly ionizes both excited states with
$m$--photon and $(m-1)$--photon absorption, respectively. Double ionization is established by velocity map imaging measurements of fragment ions in covariance \cite{cov4-paper,cov3-paper}.  The
pump and the probe pulses are phase-locked and the probe
pulse is delayed by
$\tau$, so we can write the electric field as $E_{\text{pr}}(t) =
E'_{0}(t-\tau) \cos(\omega_{0}(t-\tau) - \phi)$, where we control
the phase $\phi$ and delay $\tau$ with our
pulse-shaper. Making use of the expression for $\rho_{12}(\boldsymbol{R},t)$ given above, we can express the dication yield as:
     \begin{equation}
      Y(\boldsymbol{R},t)=|a_1|^2|b_1|^2+|a_2|^2|b_2|^2+ b_1 b_2^* \rho_{12}(\boldsymbol{R},t) + c.c.
    \end{equation}
  Here, $b_1$ and $b_2$ represent the ionization amplitudes
  for states 1 and 2, which are proportional to the $m^\mathrm{th}$ and
  $(m-1)^\mathrm{th}$ power of the electric field of the
  probe:
  \begin{subequations}
\begin{align}
    b_1(t) &= Q_{1\text{f}}\big(E_\text{pr}(t)\big)^m \\
    b_2(t) &= Q_{2\text{f}}\big(E_\text{pr}(t)\big)^{(m-1)}
    \label{eq:bn}
\end{align}    
\end{subequations} 
  where $Q_{\text{1f}}$ and $Q_{\text{2f}}$ are the matrix elements for
  ionization from state 1 or 2 respectively into the final
  dicationic state $f$. Integrating over $t$, one can arrive at an expression for the ionization yield in terms of the delay $\tau$ and
  phase $\phi$ that we control \cite{brian-long}:
    
  \begin{align}
      Y(&\boldsymbol{R},\tau, \phi) =
          |a_1|^2\,\big|Q_{1f}(E'_0)^m\big|^2+|a_2|^2\,\big|Q_{2f}(E'_0)^{m-1}\big|^2 \nonumber
    \\
    &+a_1 a_2^* Q_{\text{1f}} Q_{\text{2f}}^* (E'_0)^{(2m-1)} e^{i\phi} \rho_{12}(\boldsymbol{R},\tau) + c.c.      \label{eq:R_resolved_yield2}
  \end{align}

This expression highlights the phase dependence of the ionization yield and three mechanisms for
electronic decoherence. The first is the loss of wavefunction overlap -- the decay of the product $\chi_1(\boldsymbol{R},\tau) \chi_2^*(\boldsymbol{R},\tau)$. The
second is dephasing -- an $\boldsymbol{R}$ dependent phase advance $
\omega_{2}(\boldsymbol R)-\omega_{1}(\boldsymbol R)$, which
washes out the $\boldsymbol{R}$ integrated  yield.  The last is the loss of
population, i.e., the decay of the coefficients $a_{i}$ via internal conversion or intersystem
crossings. All three contributions to decoherence can be
suppressed for states with parallel potential energy
surfaces, which can minimize internal conversion and
dephasing, and for which the wave packet evolution is very
similar, maintaining the overlap.  Furthermore, if one is
able to perform $\boldsymbol{R}$--resolved measurements,
then one can further mitigate dephasing since one limits the
range in $\boldsymbol{R}$ over which the phase term is
integrated.  Here, we experimentally demonstrate long
lived electronic coherences between dissociative states of a
polyatomic molecule with parallel potential energy surfaces,
and illustrate how covariance velocity map imaging of the
fragment ions  arising from the dication (providing KER
resolved measurements) allows us to directly view the
entangled electronic and nuclear degrees of freedom,
which leads to dephasing and the decay of electronic
coherence.  

There have been many studies that aim to determine the dominant mechanism for the loss of coherence 
\cite{PhysRevLett.118.083001,koll2022,Despre2018,Matselyukh2022,vester2023,Dey2022,Vacher2016,csehi2020preservation}. 
While the dominant mechanism can depend on the details of the molecular system, calculations support the idea that vibrational dephasing can be mitgated in states with parallel potentials, leading to long lasting coherences \cite{vester2023}. 

In earlier work \cite{brian-coherence}, we found parallel
potentials for relatively low--lying states of the molecule
Thiophene -- states involving excitation of lone pair orbitals.  
Here we consider excitation of high lying Rydberg states, for which the
potential energy surfaces are approximately
parallel \cite{brian-coherence,gibson1991dynamics}.

In this paper we
study entangled nuclear-electronic wave packets in 1,1,1-Trifluoroacetone
(3F-Acetone). Via multi-photon excitation, we excite the
molecule to a pair of high-lying Rydberg states of the
neutral molecule. We then follow the evolution of electronic
coherences between these two states and demonstrate how
electronic coherences can be maintained even in the face of
large amplitude nuclear motion (i.e., dissociation). 

\section{Experimental Method}
\begin{figure} 
	\centering
	\includegraphics[width=0.48\textwidth]{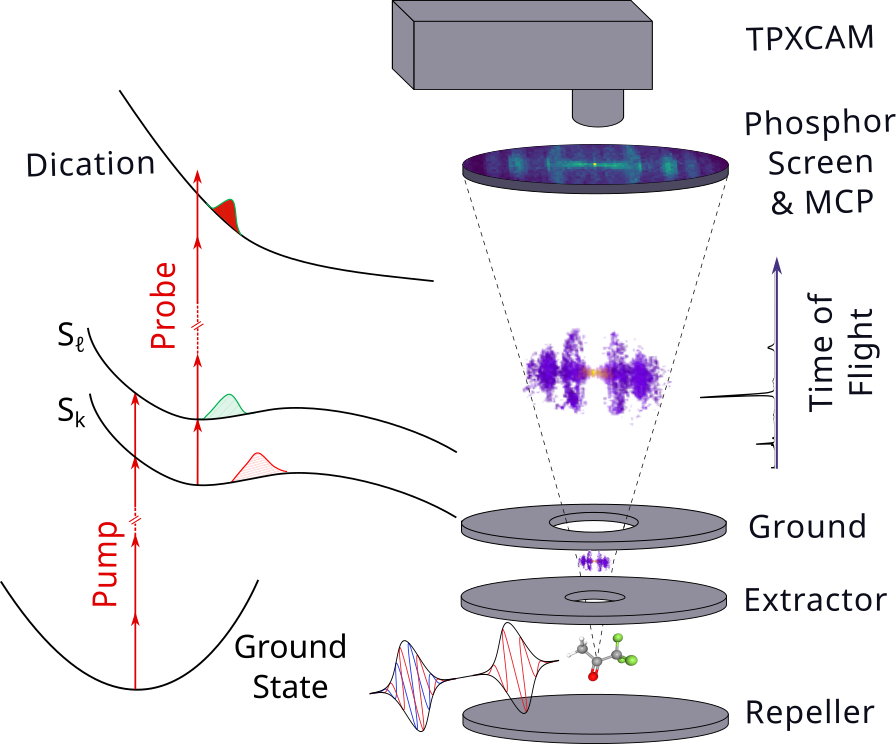}
 	\caption{Left panel: Potential energy curves and
          excitation scheme for detecting electronic
          coherences. The molecule is excited to a pair of
          high-lying Rydberg states that are separated by
          one photon energy and then doubly ionized with a
          probe pulse. Right Panel: Shaped, phaselocked pump and probe pulses with velocity map imaging apparatus for phase sensitive KER resolved measurements of fragments from molecular dications.}
 	\label{fig:apparatus}
\end{figure}

We carried out measurements using a commercial amplified
titanium sapphire laser that produces 30--fs pulses at
1 kHZ with up to 1--mJ pulse energy. The 30--fs pulses are spectrally
broadened using an Argon-filled 2.1--m long hollow--core fiber with inner diameter
of 450 $\mu$m (see \autoref{fig:spectrum-cfrog} in the
Appendix for the spectrum).  The broadened pulses are then
compressed and shaped using an acousto-optic modulator
(AOM) pulse shaper \cite{aom-paper}.

The initial electric field 
$E(\nu)$ is shaped in the frequency domain (at the Fourier plane of the pulse shaper) by an acoustic mask $M(\nu)$ programmed onto the AOM to produce a pump-probe pulse pair. 
The shaped electric field $\widetilde E(\nu)$ is given by
 $\widetilde E(\nu) = M(\nu) E(\nu)$. For control of the pump-probe delay and relative phase
between the pulses, we choose a mask of the form

\begin{equation}
  M(\nu) = A_{tot}\Big(1+ A_R
  \exp\big(i 2\pi\tau(\nu-\nu_{L})  + i \phi_L\big)\Big)
\end{equation}
where $A_{tot}$ controls the overall amplitude, $R$ the relative amplitude ($E'_0=A_RE_0)$, $\phi_L$ the relative phase and $\tau$ the relative delay between the two pulses. The quantity $\nu_{L}$ is the locking frequency and writes a delay dependent relative phase such that there is always constructive interference at the frequency $\nu_{L}$ (for $\phi_L=0$), regardless of the
delay between the two pulses. Thus the controllable phase of the laser can be described by $\phi=\phi_L-2\pi\nu_L\tau$.

In addition to control of the pulse pair, the
pulse shaper also allows us to characterize the pulses we
generate. We perform pulse-shaper-assisted,
second harmonic generation collinear frequency resolved optical gating (PS-CFROG) to determine the temporal profile of
our pulses \cite{aom-paper}. The resulting compressed and shaped pulses have a minimum duration of
7 fs FWHM (see \autoref{fig:spectrum-cfrog} in the appendix). We
made extensive use of the AOM pulse shaping capability to
compress, control, and characterize the laser pulses \cite{aom-paper}.

The shaped pulses are focused onto an effusive molecular beam of
3F-Acetone inside the velocity map imaging
spectrometer using a 5 cm focal length 
concave silver mirror, which is inside the vacuum chamber. The base
pressure of the vacuum chamber is $10^{-10}$ Torr. The molecule 3F-Acetone is introduced into a separate
sample chamber, and then skimmed into the reaction chamber
in order to yield an effusive beam. This effusive beam
overlaps with the laser focus at the center of our VMI
apparatus. The partial pressure of 3F-Acetone inside the
reaction chamber was kept around $5\times 10^{-8}$ Torr in
order to yield the desired count rate of around $5-10$ ion
hits per laser shot.

We collect ion momentum data
using a velocity map imaging system, which has funnel micro channel
plates with open area of $\sim$90\% and a P47 phosphor screen
(depicted schematically in \autoref{fig:apparatus}). The
hits are recorded by a Timepix camera, whose pixels have a
time of arrival resolution of 1.5 ns
\cite{Nomerotski2019,PhysRevA.102.052813}, allowing us to simultaneously record and resolve all of the fragment ions produced in a given laser shot. Furthermore, the Timepix camera 
allows us to reconstruct the 3D momentum of ions using the time
of flight of ions and their position ($[x,y,t]\mapsto [p_x,p_y,p_z]$) without the need for Abel
inversion \cite{original-vmi-paper,PhysRevA.102.052813}. We 
estimate the peak laser intensity for an unshaped laser pulse at the focus to be 550 TW/cm$^2$ (corresponding to individual pump and probe intensities of about 230 TW/cm$^2$ and 550 TW/cm$^2$, respectively), which we have obtained from the
$2U_p$ cutoff of above-threshold ionization of Argon.


\section{Results}

\begin{figure}[t]
  \centering
  \includegraphics[width=.49\textwidth]{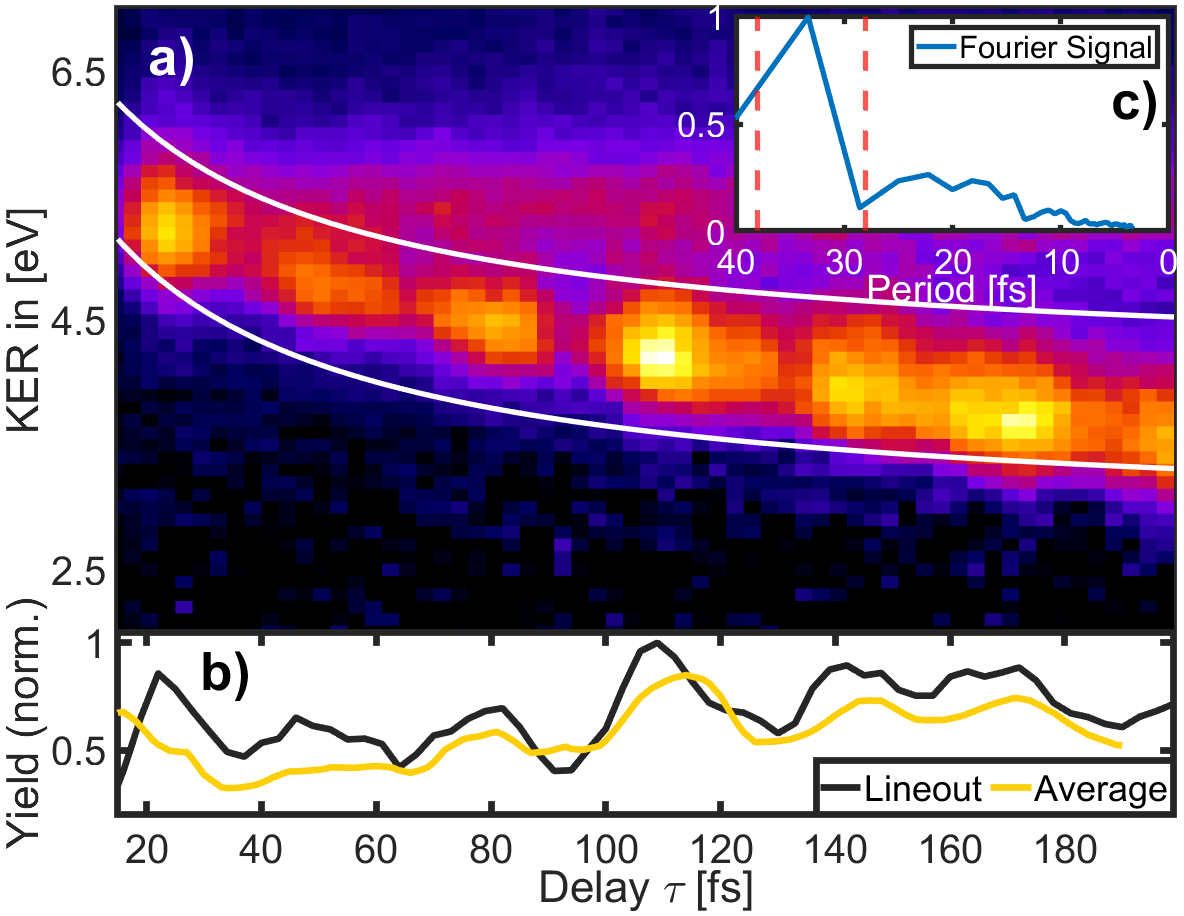}
  \caption{a) Kinetic energy release (KER) resolved
  covariance between \cf\ and \co\ ions as a function of pump-probe delay. The $y$
  axis shows the total KER from these
  ions and the $x$ axis represents the pump-probe
  delay. b) is the yield integrated
   between the two white lines (black) as well as the yield averaged over all locking frequencies (yellow). Insert: Fourier analysis of the locking frequency averaged yield}
\label{fig:pp-cf3-cov-ch3co-ker-cropped}
\end{figure}

\begin{figure*}[t]
	\centering
	\includegraphics[width=1\textwidth]{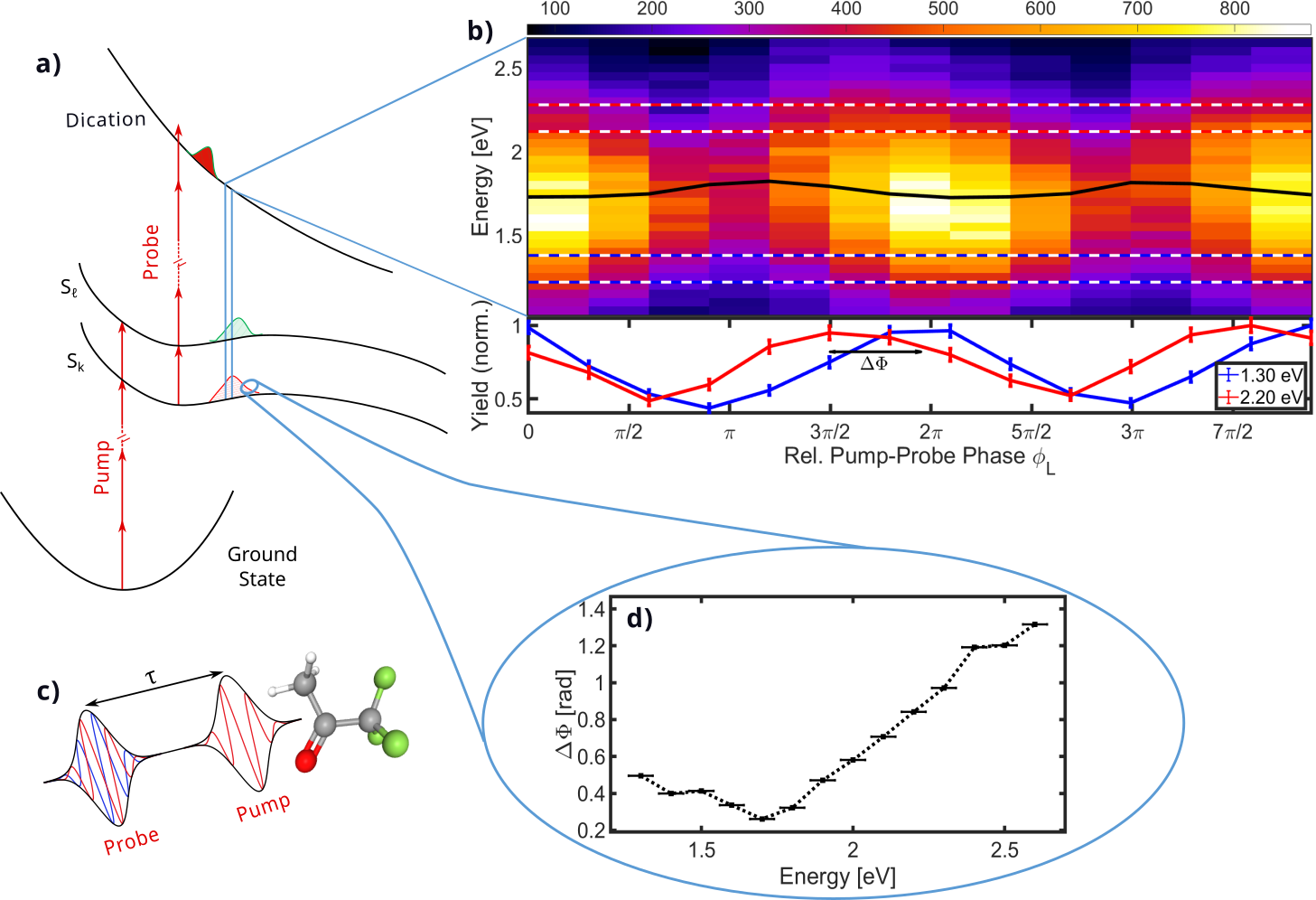}
 	\caption{Measurement scheme with \cf\ energy and
          phase dependent yield. a) potential
          energy curves for states relevant to the creation
          and detection of a coupled electron-nuclear wave
          packet.  b) \cf\ double
          ionization yield as a function of \cf\ energy and pump-probe delay
          with lineouts below for two different energy
          regions as well as the center of mass of the
          distribution in black. 
          c) Schematic of shaped, phaselocked pump  and probe pulses with different relative phases
 d) Phase extracted from fitting the data in
  panel (b) as a function of \cf\ energy.  }
 	\label{fig:fancy-cartoon}
\end{figure*}

We measured the momenta of fragment ions from 3F-Acetone in
covariance as a function of the delay and the relative phase
between the two pulses. We focus on the fragmentation of the
molecule along the \cfn--\con\ bond using covariance velocity
map imaging. In particular, we look at the correlation of
\cf\ and \co\ ions in the dataset on a shot by shot
basis, which allows us to hone in on the dissociation along
the \cfn--\con\ coordinate.  Working in covariance provides the same information as coincidence, but allows us to work in a
higher count rate regime
\cite{cov4-paper,cheng2022strong,crati-paper,cov3-paper}.

In \autoref{fig:pp-cf3-cov-ch3co-ker-cropped} we present the
results of the pump-probe measurements, which show the
energy resolved covariance yield of \cf\ and \co\ ions as a
function of the pump-probe delay. Measuring these two
fragment ions in covariance and checking for momentum
conservation confirms that they come from the same molecular dication, and allows us to draw some simple conclusions about the dynamics between pump and probe pulses.  There are two prominent
features in \autoref{fig:pp-cf3-cov-ch3co-ker-cropped}: the kinetic
energy release decreases with pump--probe delay, and the
covariance yield is modulated with a period of
$33\pm5$fs. 
The monotonic decrease in kinetic energy release with
pump--probe delay is consistent with dissociation along
this bond, while the modulations in the yield are consistent
with vibrations during dissociation.

As the KER resolved pump--probe measurements of the fragment
ions point toward dissociative dynamics, we considered
electronic states of both the neutral and monocation.  The
ground state of the monocation is dissociative along the
\co --CF$_3$ coordinate
\cite{cardoza2005understanding,cardoza2005interpreting},
meaning that the dynamics could take place on low lying
states of the cation or high lying Rydberg states of the
neutral, which are parallel to the lowest cationic state.  We
argue below based on phase dependent measurements that it is
unlikely to be cationic states. 


In order to interpret the modulations in the covariance
yield, we calculated the vibrational frequencies of the
neutral ground state at the B3LYP level of theory.  The
modes with frequencies near the measured modulation
frequency are listed in (\autoref{tab:vibrations}) of the
appendix. Given that we expect the excited states of the
molecule to have slightly lower frequencies and that the
modulations have a pronounced effect on the CF$_3^+$ and
\co\ yield, we suspect that it is \cfn\ vibrations which are underlying the modulations in the measurements.   In
\autoref{fig:vibrational-motion} in the appendix
we depict the motion of the anti-symmetric \cfn\ mode that might
give rise to the modulations we see in \autoref{fig:pp-cf3-cov-ch3co-ker-cropped}. 

With the pulse shaper, we have the ability to perform
pump--probe measurements with independent control over the phase
and delay between pump and probe pulses.  This allows us to
confirm that the modulations in the covariance yield arise
from a vibrational coherence rather than an electronic one.
The bottom panel of
\autoref{fig:pp-cf3-cov-ch3co-ker-cropped} compares the
measurements carried out for a fixed phase between pump and
probe pulses with measurements that are averaged over all phases
at each delay.  Both measurements show clear modulations in
the yield, consistent with these arising from vibrations
which modulate the ionization yield.  

While pump--probe measurements which average over phase allow us to focus on vibrational dynamics, phase scans at fixed delays highlight electronic coehrences, as shown in \autoref{eq:R_resolved_yield2}. In
\autoref{fig:fancy-cartoon} we present the phase scan at a fixed
delay of 95 fs. We chose this delay partially to rule out any optical interference between the pump
and the probe pulses, and partially to highlight the long lived electronic coherences which persist during dissociation. In the appendix we detail how we have
ruled out optical interference of the
two pulses. Since we see a single modulation of the energy
resolved yield in \autoref{fig:fancy-cartoon} within 
2$\pi$ phase, as opposed to multiple modulations within 2$\pi$, the
states that contribute to the electronic
coherence must be separated by a single photon
energy \footnote{In general, if the states were separated by
  $K$ photons we would see $K$ modulations within $2\pi$ as
  the phase term in \autoref{eq:R_resolved_yield2} would
  become $e^{iK\phi}$}.
  
Single point electronic structure 
calculations at the
Frank--Condon point for the ion show that there are
no pairs of states of the cation that are separated by a
single photon. We thus conclude that the electronic coherence
is coming from high--lying states of the neutral
molecule. The fact that the states are high lying also has
the consequence that the potential energy surfaces are likely
close to being parallel to each other \cite{brian-coherence,gibson1991dynamics}. This effect decreases the
decoherence of the two states due to loss of wave function
overlap as outlined in the introduction.

\autoref{fig:fancy-cartoon} shows an explicit manifestation of
entanglement between nuclear and electronic degrees of
freedom. Here we plot the energy
resolved yield of the \cf\ fragment, instead of the covariant kinetic
energy release from \cf\ and \co\ 
 since the statistics are better for the single fragment
 yield. We have, however, used the
covariance method to confirm that the plotted yield is
indeed coming from the dissociation along the \cfn--\con\
bond.  Note that the kinetic energy of the \cf\
fragment is determined by the \cfn--\con\ distance. As the kinetic energy of the \cf\ changes, so
does the position of constructive interference of the two
wave packets. This is highlighted by the variation in the first moment of the energy distribution, which is plotted in black on top of the 2D color plot. We make this connection explicit by plotting in panel d)
the phase shift, $\Delta \Phi$, between the different lineouts (black arrow shown with the red and blue curves in the bottom of panel b))
from \autoref{fig:fancy-cartoon} as a function of kinetic energy of the
\cf\ fragment. 

\begin{figure}[t]
  \centering
  \includegraphics[width=.49\textwidth]{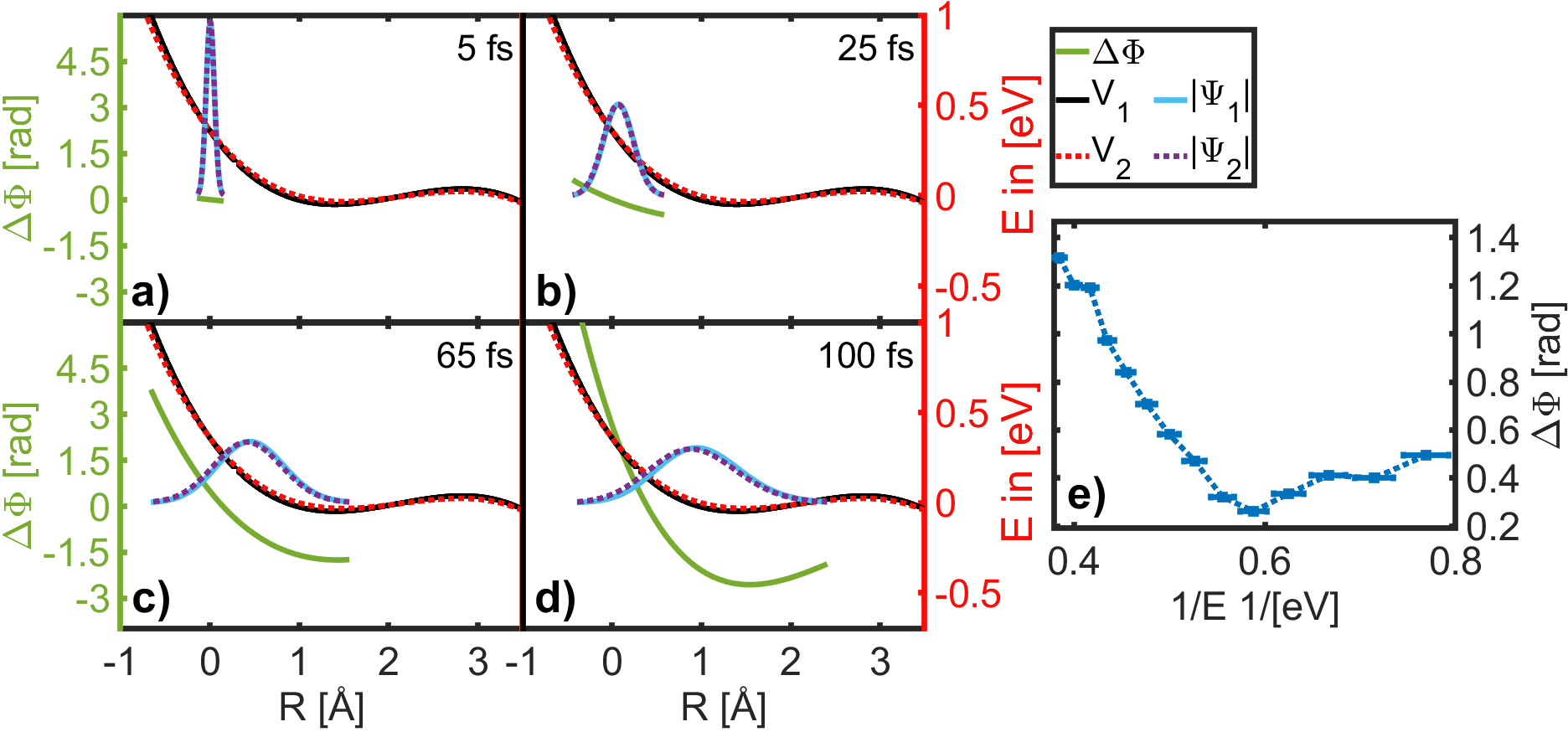}
  \caption{a)--d): Time evolution of two wave packets
    on slightly different dissociative
    potentials. In green is the $R$-dependent phase difference between the
    wavefunctions. The wave functions are scaled for easier viewing. e) Phase extracted from fitting the energy dependent
  modulations of the phase scan (see also (d) panel of
  \autoref{fig:fancy-cartoon}). The $x$-axis is converted
  into $1/E$ to represent $R$ because of the Coloumb
  potential relation $E \sim 1/R$.
}
\label{fig:tdse}
\end{figure} 

As a test of the basic idea that our measurements are
sensitive to the KER dependent (and thus the $\boldsymbol R$ coordinate dependent) phase difference between the wave functions on the two electronic states, we carried out   calculations where we solve the time dependent Schr\"odinger equation (TDSE) for a model 1D potential.   We consider two wave packets propagating on
slightly different parallel potentials. We employ the split
operator formalism to propagate the initial wave functions
on the two dissociative potentials and plot the
evolution of the phase difference between the two
wave-functions. \autoref{fig:tdse} shows the results of the calculations.   We see a qualitative agreement between the
measured phase difference and the phase difference of two
wave-functions. The potential used in the calculations is based on our expectation for the high lying Rydberg states of the molecule, which are parallel to the ground state of the molecular cation \cite{gibson1991dynamics,cardoza2005understanding}.  We defined the potential in the calculations as a cubic spline
between four points, roughly setting the position of the initial
roll down, the minimum, the barrier height and the roll down
after the barrier. We fixed these four points and generated a series of
potentials by randomly varying around these
points in order to determine the sensitivity of the
calculations to the details of the potential energy curve on
which the wave packets evolve.  Our calculations showed the
same qualitative features independent of the exact shape of the
potential: As shown in panel (a), the wave packets initially
have very little phase 
difference because they haven't propagated on the
potentials for enough time to develop a phase difference. As
the wave packets pick up momentum on slightly different
potentials, we see a linear phase difference between them
(b). As the wave packets further progress down the potential and experience less acceleration or are
slowed down by the barrier, the linear
phase difference evolves into a hockey stick shaped feature (c)--(d). We
have confirmed that this hockey stick
shaped feature in panel (c) persists qualitatively as we vary the
potentials around the fixed points. This furthers our interpretation that the
measurements we show in \autoref{fig:fancy-cartoon} are
indeed resulting from the coherent interference of two
wave-packets  propagating on approximately
parallel dissociative potential energy surfaces.

\section{Conclusion}
In conclusion, we have presented measurements of entangled
nuclear--electronic wave packet dynamics in a multimode molecular
system. In particular, we have observed a pump--probe phase--dependence of the energy resolved yield of molecular fragments (\autoref{fig:fancy-cartoon}).  A phase locked pump--probe pulse pair launches and
interrogates the wave packet dynamics via multiphoton excitation and
ionization.  Fragment ions are measured with momentum
resolved covariance velocity map imaging. In the pump--probe scan we observe a dissociation along the \cfn--\con\ bond which is coupled with vibrational excitation of the molecule (\autoref{fig:pp-cf3-cov-ch3co-ker-cropped}).  Our measurements highlight the
entanglement between electronic and nuclear degrees of
freedom, and demonstrate the maintenance of electronic
coherence despite large amplitude nuclear motion. 

\section*{Acknowledgements}

G.M., B.K., C. C. and T.C.W. were supported by the National Science Foundation
under award number 2110376.  I. B-I was supported by the Chemical Sciences, Geosciences, and Biosciences Division, Office of Basic Energy Science, Office of Science, U.S. Department of Energy, under Award No. DE-FG02-86ER13491.

\section*{Conflict of Interest}

The authors declare that they have no conflicts of interest.

\section*{Data Availability}

\customlabel{sec:data}{Data Availability}

The data that support the findings of this study are available from the corresponding author upon reasonable request.

\section*{Appendix}
\begin{figure}[h]
   \centering
   \includegraphics[width=0.49\textwidth]{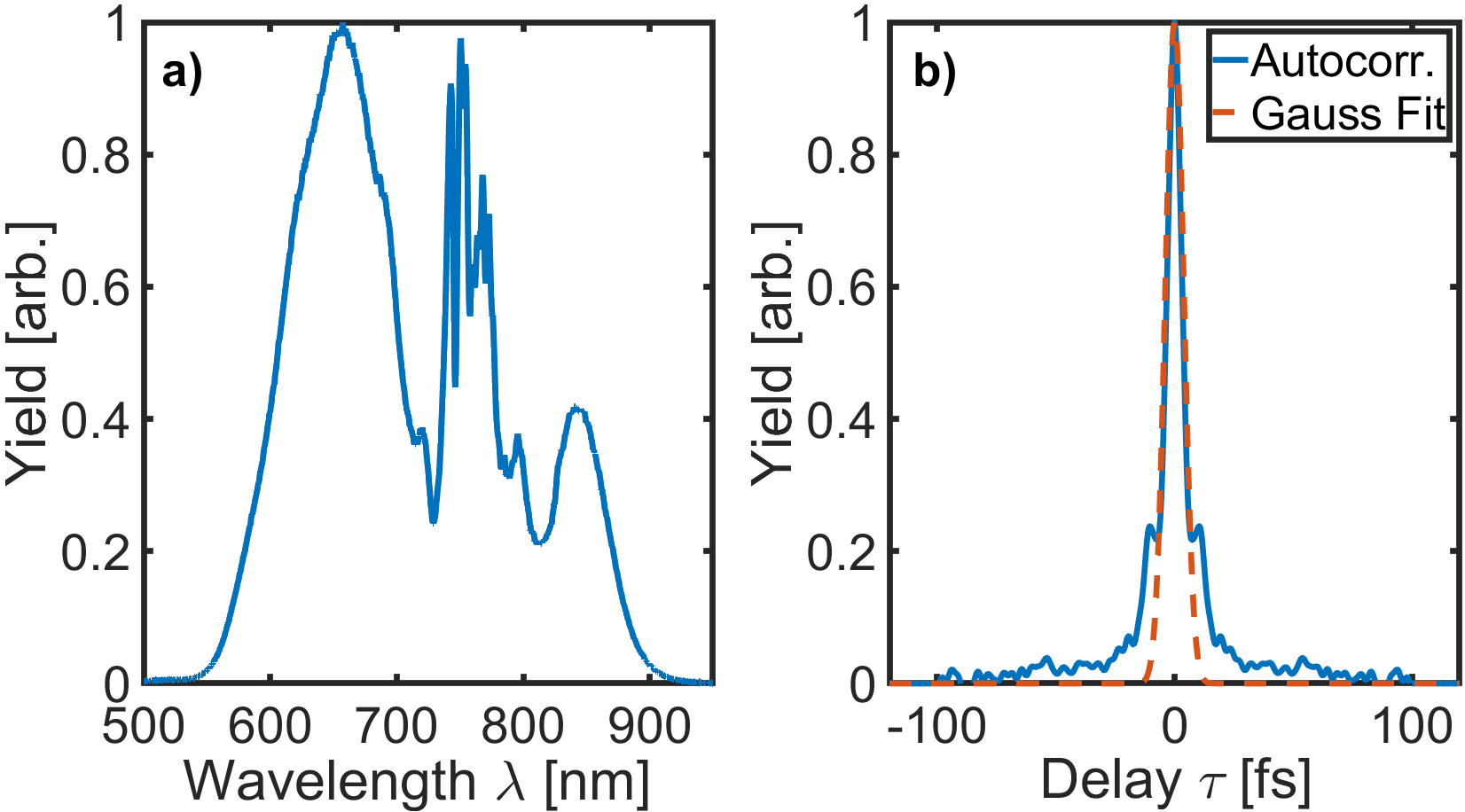}
   \caption{a) Spectrum of the laser pulse out of the
     hollow--core fiber. The spectrum is structured,
     consistent with self--phase modulation in hollow-core
     fibers. b) Intensity cross correlation signal from
     CFROG and the Gaussian fit to it. Note that the
     $\mathrm{FWHM}$ is 7 fs and there are no pre-- or post--pulses at large delays.}
   \label{fig:spectrum-cfrog}
 \end{figure}

We took care in ruling out any optical interference
between the pump and the probe pulses. We checked for
optical interference in two independent ways. The first was to perform detailed pulse characterization measurements, using PS-CFROG to characterize the pulses.   
In \autoref{fig:spectrum-cfrog} we plot the spectrum and the CFROG
of the compressed pulse. Note that while there is some
structure in the pulse around 10 fs, there are no pre or
post pulses at longer ($\sim$ 95 fs) delays where we carry out our
interference measurements. 

\begin{figure}[h!]
   \centering
   \includegraphics[width=0.49\textwidth]{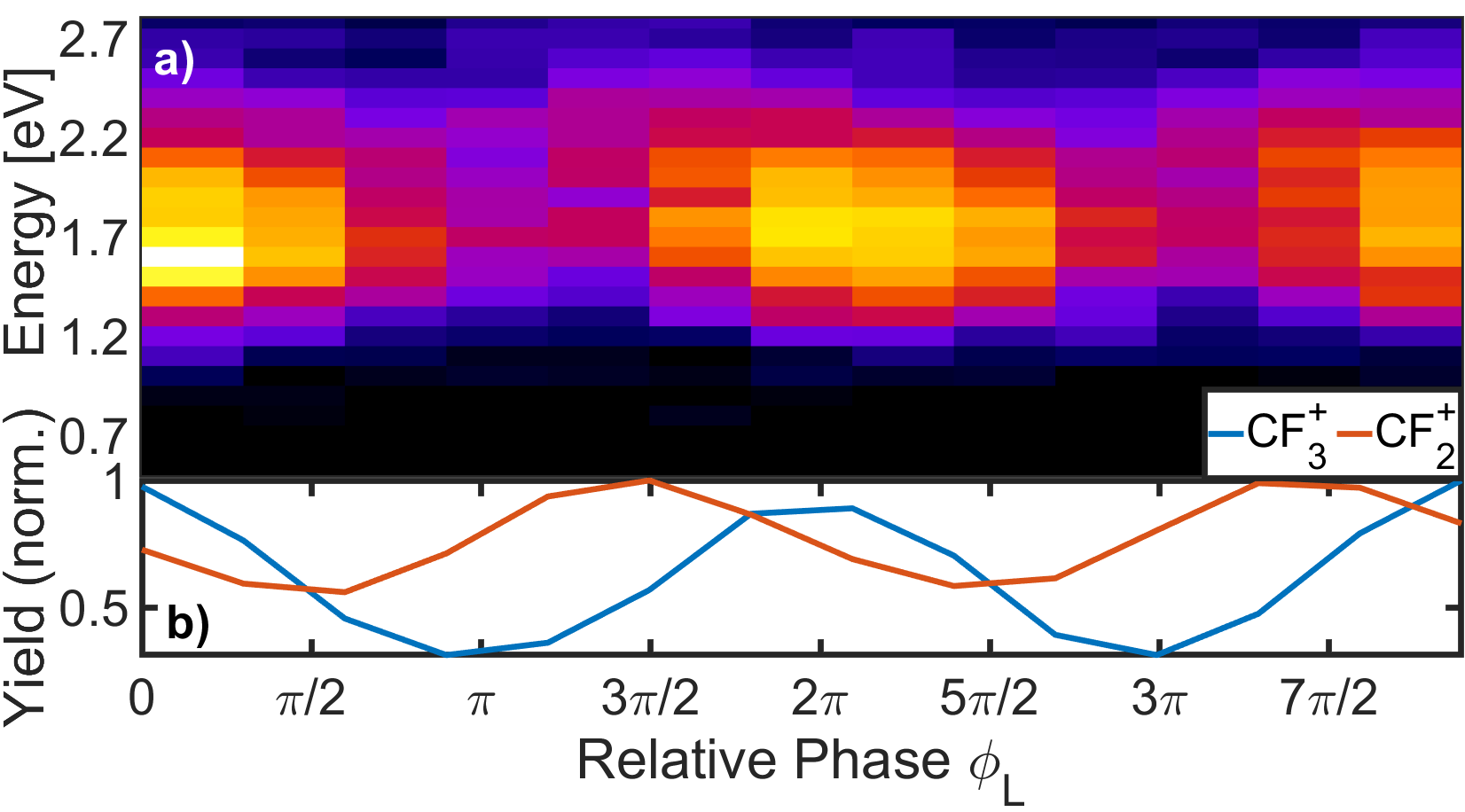}
   \caption{a) Energy Resolved Yield of \cf\ in covariance
   with \co\ ion. b) Marginal of the above
   covariance yield (blue) and marginal of the yield
   of CF$_{2}^{+}$}
   \label{fig:phase-95fs-cf3-cov-ch3co-yield-cf2-cropped}
 \end{figure}

Our second independent check is looking at the phase dependence of different fragment ions. If optical interference were giving rise to the
modulations seen in \autoref{fig:fancy-cartoon} it
would be because the total electric field varies with phase
as a result of constructive and destructive interference.
In this case the phase dependence of the yields of all fragments in a
given dataset would be the same because the modulations are
only due to variation in the total intensity. This is,
however, not the case as we show in
\autoref{fig:phase-95fs-cf3-cov-ch3co-yield-cf2-cropped}. There
we can  see that the 
\cf\ and CF$_{2}^{+}$ yields are phase shifted relative to
each other. Combined with our
characterization of the pulses, we conclude that the phase
variation is due to dynamics in the molecule and is not an
optical interference effect.

\vfill

\vfill

 \begin{table}[h!]
   \centering
   \begin{tabular}{c|c|c}

     Assignment \cite{Durig1980VibrationalSA} & Frequency [cm$^{-1}$] & Period [fs] \\
     \hline
                CH$_{3}$ symmetric rock&962& 35\\
                CH$_{3}$ antisymmetric rock&1027&32 \\
                CF$_{3}$ symmetric stretch&1131& 29\\
               \textbf{ CF$\boldsymbol{_{3}}$ antisymmetric stretch}&\textbf{1189}&\textbf{28} \\

   \end{tabular}
   \caption{Various vibrational exitations of 3F-Acetone
     which have a period around $33\pm5$ fs at the
     Frank-Condon point of the ground state calculated at
     B3LYP level of theory }
   \label{tab:vibrations}
 \end{table}
 \vfill

 \begin{figure}[h!]
   \centering
   \includegraphics[width=0.49\textwidth]{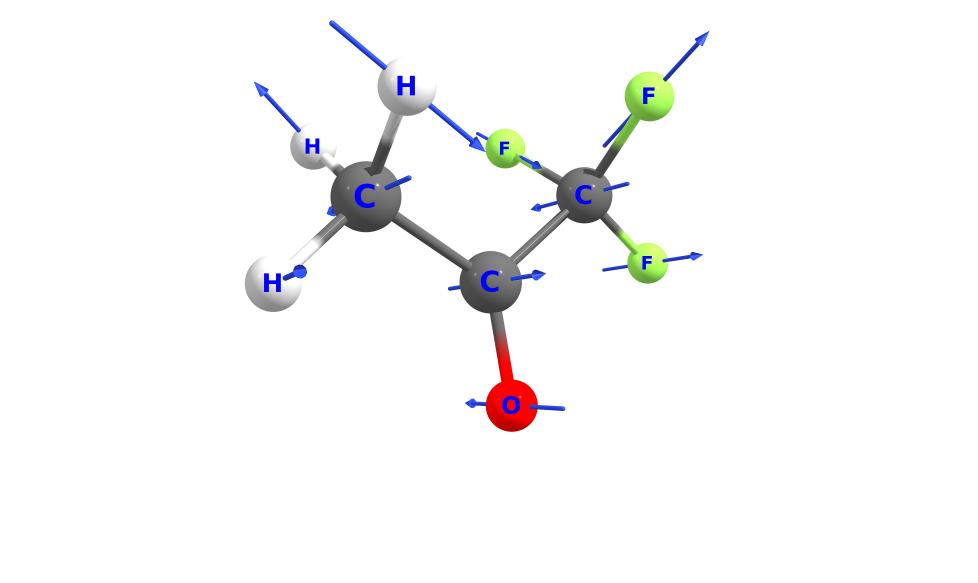}
   \caption{Displacement vectors for antisymmetric stretch
     of \cfn }
     \label{fig:vibrational-motion}
 \end{figure}  
 \vfill
\clearpage

\bibliographystyle{apsrev4-2}
\bibliography{3f-acetone}



\vfill

\end{document}